\documentclass[10pt,onecolumn]{IEEEtran}
\usepackage{fullpage}
\usepackage{dsfont} 
\usepackage{xcolor}
\usepackage{authblk}
\PassOptionsToPackage{dvipsnames}{xcolor}
\definecolor{red}{HTML}{E51400}  
\definecolor{blue}{HTML}{0050EF} 
\definecolor{green}{HTML}{008A00} 
\definecolor{purple}{HTML}{AA00FF} 
\definecolor{dark-red}{rgb}{0.4, 0.15, 0.15}
\definecolor{dark-blue}{rgb}{0.15, 0.15, 0.4}
\definecolor{medium-red}{rgb}{0.5, 0, 0}
\definecolor{medium-blue}{rgb}{0, 0, 0.5}
\definecolor{light-red}{rgb}{0.7, 0, 0}
\definecolor{light-blue}{rgb}{0, 0, 0.7}
\definecolor{bv}{RGB}{50, 74, 178}
\definecolor{rb}{RGB}{230, 43, 30}

\usepackage{array}
\usepackage{setspace}
\usepackage[
    colorlinks,
    linkcolor={purple},
    citecolor={blue},
    urlcolor={red},
    bookmarks=true,
    linktoc=page
    ]{hyperref}

\usepackage{fancybox,framed}
\usepackage{nicefrac}
\usepackage{booktabs}
\usepackage{nicematrix}
\usepackage{multirow,multicol}
\usepackage{cite}
\usepackage{xcolor} 
\usepackage{enumitem}
\setlist[itemize]{label=-}

\usepackage{amsmath}
\usepackage{amssymb}
\usepackage{amsthm}
\usepackage{thmtools}
\usepackage{physics}
\usepackage{booktabs}
\usepackage{hyperref}
\usepackage[square,sort,comma,numbers]{natbib}
\theoremstyle{plain}

\newtheorem{theorem}{Theorem}
\newtheorem{lemma}[theorem]{Lemma}

\theoremstyle{definition}
\newtheorem{definition}{Definition}

\theoremstyle{remark}

\newcommand{\cM}{\mathcal{M}}
\newcommand{\cF}{\mathcal{F}}
\newcommand{\cK}{\mathcal{K}}
\newcommand{\cX}{\mathcal{X}}
\newcommand{\csk}{C_{\mathrm{SK}}}
\newcommand{\cP}{\mathcal{P}}
\newcommand{\cS}{\mathcal{S}}
\newcommand{\cE}{\mathcal{E}}
\newcommand{\cH}{\mathcal{H}}
\newcommand{\unif}{\text{unif}}
\newcommand{\spanv}{\text{span}}

\usepackage[linesnumbered,ruled,vlined]{algorithm2e}

\usepackage{tikz}
\usepackage{subcaption}
\usetikzlibrary{arrows.meta, positioning}

\SetKwInput{KwInit}{Init}

\tikzset{
  copyone/.style={-, thick, blue},
  copytwo/.style={-, thick, red}
}

\def\BibTeX{{\rm B\kern-.05em{\sc i\kern-.025em b}\kern-.08em
    T\kern-.1667em\lower.7ex\hbox{E}\kern-.125emX}}
\title{Perfect Secret Key Generation\\ for a Class of Hypergraphical Sources}

\author{
\IEEEauthorblockN{Manuj Mukherjee$^\dag$} \hspace {1cm} \and \IEEEauthorblockN{Sagnik Chatterjee$^\ddag$}\hspace {1cm} \and \IEEEauthorblockN{Alhad Sethi$^*$}  
}

\begin{document}

\maketitle

\renewcommand{\thefootnote}{}
\footnotetext{
$^\dag$ M. Mukherjee is with the Indraprastha Institute of Information Technology Delhi, Delhi, India. Email: manuj@iiitd.ac.in.\\

$^\ddag$ S. Chatterjee is with the Tata Institute of Fundamental Research, Mumbai, India. Email: chatsagnik@gmail.com.\\

$^*$ A. Sethi is with the Indian Institute of Science, Bangalore, India. Email: alhadsethi@iisc.ac.in.\\

The work of S.~Chatterjee was funded by the Department of Atomic Energy, Government  of India, under project no. RTI4014.
}

\renewcommand{\thefootnote}{\arabic{footnote}}

\begin{abstract}
Nitinawarat and Narayan proposed a perfect secret key generation scheme for the so-called \emph{pairwise independent network (PIN) model} by exploiting the combinatorial properties of the underlying graph, namely the spanning tree packing rate. This work considers a generalization of the PIN model where the underlying graph is replaced with a hypergraph, and makes progress towards designing similar perfect secret key generation schemes by exploiting the combinatorial properties of the hypergraph. 

Our contributions are two-fold. We first provide a capacity achieving scheme for a complete $t$-uniform hypergraph on $m$ vertices by leveraging a packing of the complete $t$-uniform hypergraphs by what we refer to as star hypergraphs, and designing a scheme that gives $\binom{m-2}{t-2}$ bits of perfect secret key per star graph. Our second contribution is a 2-bit perfect secret key generation scheme for 3-uniform star hypergraphs whose projections are cycles. This scheme is then extended to a perfect secret key generation scheme for generic 3-uniform hypergraphs by exploiting star graph packing of 3-uniform hypergraphs and Hamiltonian packings of graphs. The scheme is then shown to be capacity achieving for certain classes of hypergraphs. 
\end{abstract}

\section{Introduction}\label{sec:intro}

The multiparty secret key generation problem considers $m$ parties that need to agree upon a group secret key. To facilitate secret key generation, the parties are given access to a \emph{discrete memoryless multiterminal source (DMMS)}, i.e., i.i.d. copies of distinct but correlated random variables, and are allowed to communicate interactively over a noiseless but public broadcast channel. The secret key generated, therefore, needs to be secured from any passive eavesdropper\footnote{An eavesdropper is said to be passive if it can only listen to the messages but not tamper with them.} listening to the communication sent over the channel. Initially posed for $m=2$ parties independently by \cite{Maurer93, AC93}, this problem was later extended to an arbitrary number of parties by the seminal work of Csisz\'{a}r and Narayan \cite{CN04}. 

The key quantity of interest in these works is the \emph{secret key capacity}, i.e., the maximum \emph{rate} of a secret key that can be generated, a full characterization of which appears in \cite{CN04}. An alternate characterization of the secret key capacity was given by Chan and Zheng \cite{CZ10}, who showed that the secret key capacity equals a multivariate analogue of the mutual information \cite{chan15}, which we shall refer to as \emph{multipartite information}.

Subsequent works have looked into other aspects of the multiparty secret key generation problem, such as communication rate needed to achieve secret key capacity \cite{tyagi, mukherjee,ccde,rskmax,hypupper,pinedmond,ZC20}, capacity with silent parties \cite{AG,mukherjee,zhang,so}, `single shot' secret key capacity \cite{tw}, and \emph{perfect} secret key generation \cite{nitinrawat,chanlin,ccde}. In this work, we focus on the aspect of perfect secret key generation. 

A \emph{perfect secret key} satisfies the more stringent condition of being completely independent of the communication, as opposed to the usual notion of a \emph{strong secret key} where the key is allowed to be `almost independent' of the communication -- see Section~\ref{sec:model} for a precise definition. While a general characterisation remains unknown, perfect secret key capacity is known to match the usual (strong) secret key capacity for certain classes of sources such as the \emph{pairwise independent network (PIN)} model \cite{pinorig}. 

The PIN model is defined using an underlying graph whose vertex set equals the set of parties. Associated with every edge are i.i.d. sequences of independent Be$(1/2)$ random variables, and each party has access to the random variables associated with the edges incident with it. Nitanawarat and Narayan \cite{nitinrawat} gave an elegant capacity achieving perfect secret key generation scheme for the PIN model. The scheme involves packing multiple copies of the underlying graph with spanning trees and generating a single bit of perfect secret key from every spanning tree. This scheme therefore achieves a perfect secret key of rate equal to the spanning tree packing rate of the graph, which is shown to equal the (strong) secret key capacity by exploiting standard results on spanning tree packing such as Tutte's Theorem~\cite[Theorem~2.4.1]{diestel}. 

A natural generalization of the PIN model is the so-called \emph{hypergraphical source} \cite{ccde} where the underlying graph is allowed to be a hypergraph. One envisages that a Nitinawarat-Narayan type scheme exploiting the combinatorial properties of the underlying hypergraph should be able to generate capacity achieving perfect secret keys. In this work, we make progress towards realizing this goal.

Our contributions in this work are two-fold. Firstly, we provide an explicit secret key achieving scheme for the complete $t$-uniform hypergraph by exploiting its combinatorial properties. We begin by developing a key generation protocol that extracts $\binom{m-2}{t-2}$ bits of perfect secret key from \emph{star hypergraphs} obtained from complete $t$-uniform hypergraphs -- see Section~\ref{sec:tunif} for formal definitions. Therefore, any packing of the complete $t$-uniform hypergraph by such star hypergraphs leads to a perfect secret key of rate equal to the packing rate times $\binom{m-2}{t-2}$. The scheme is then shown to match the secret key capacity by constructing a packing of sufficient rate. We remark here that our scheme is closely related to the common randomness generation scheme for hypergraphical sources appearing in \cite{HTKZPW}.  

Secondly, we give a perfect secret key generation scheme (not necessarily capacity achieving) for generic $3$-uniform hypergraphs. The key building block of this scheme consists of a 2-bit perfect secret key generation scheme for star hypergraphs whose projected graph is a cycle. This is then exploited to show that any star hypergraph generates a perfect secret key of size two times the \emph{Hamiltonian packing number} of its projected graph. The final scheme then involves packing the original hypergraphs with star hypergraphs whose projected graphs have large Hamiltonian packing numbers. This scheme is shown to be capacity achieving for certain classes of 3-uniform hypergraphs. 

The remainder of the paper is organised as follows. Section~\ref{sec:model} introduces the problem setup, followed by Section~\ref{sec:hyper} which details the key object of our study, the hypergraphical source model. The capacity achieving perfect secret key generation scheme for complete $t$-uniform hypergraphs appears in Section~\ref{sec:tunif}. A perfect secret key generating scheme for $3$-uniform hypergraphs appears in Section~\ref{sec:threeunif}. The paper concludes with Section~\ref{sec:conc}. Proofs of certain technical results are relegated to the Appendices for better readability.

\section{Technical preliminaries}

\subsection{Multiterminal source model}\label{sec:model}

The multiterminal source model of Csisz\'{a}r and Narayan \cite{CN04} considers a set $\mathcal M = \{1,2,\dots,m\}$ of $m\geq 2$ parties. 
Party $i \in \mathcal M$ observes an $n$-length random vector $X_i^n$ taking values in a finite alphabet
$\mathcal X_i^n$.
The collection of observations across all parties is denoted by $X_{\mathcal{M}}^n = (X_i^n : i \in \mathcal{M})$. We assume that the source is memoryless, meaning that the joint distribution factors as $P_{X_{\mathcal{M}}^n}(x_{\mathcal{M}}^n) = \prod_{t=1}^n P_{X_{\mathcal{M}}}(x_{\mathcal{M},t})$, where $P_{X_{\mathcal{M}}}$ is a joint probability mass function on $\prod_{i \in \mathcal{M}} \mathcal{X}_i$. For any subset $A \subseteq \mathcal{M}$, we define the notation $X_A = (X_i : i \in A)$.

To generate secret keys, the parties are allowed to (interactively) communicate over a noiseless public channel. The random variable corresponding to the public communication is denoted by
$F := (F_1,\dots,F_r)$, where $F_j$ denotes the $j$th transmission taking values in some finite set $\cF_j$. The set of all possible values of the public communication is denoted by $\cF=\prod_{i\in r}\cF_r$. The communication being interactive, the $j$th transmission, sent by say party $i$, is a function of party $i$'s local observation $X_i^n$ as well as past public messages $F^{j-1}$. Following the public messages each party $i \in \mathcal M$ computes a local key $K_i = \varphi_i(X_i^n, F)$ using a deterministic mapping $\varphi_i:\cX_i^n\times\cF\to\cK$, where the finite set $\cK$ is referred to as the key alphabet.\footnote{Observe that the random variables $F,K_i$ and the sets $\cF,\cK$ implicitly depend on $n$.}

\begin{definition}
A number $R \ge 0$ is an \emph{achievable (strong) secret key (SK) rate} if, for every $\epsilon > 0$ and for all sufficiently large $n$, there exist a public communication $F$, local keys $K_i, i\in\cM$ and a random variable $K$ (referred to as the \emph{secret key}) taking values in the alphabet $\mathcal{K}$ such that the following conditions hold:
\begin{enumerate}
    \item \textbf{Recoverability:} For each party $i \in \mathcal{M}$, the local key $K_i$ computed from $(X_i^n, F)$ satisfies $\Pr(K_i = K) \ge 1 - \epsilon$.
    \item \textbf{Strong Secrecy:} The secret key $K$ satisfies $I(K; F) \le \epsilon$.
    \item \textbf{Almost uniformity}: The secret key $K$ satisfies $\frac{1}{n}H(K)\geq \frac{1}{n}\log\mid\cK\mid-\epsilon$.\footnote{Throughout this paper, logarithms are taken base $2$, and all entropy quantities are measured in bits.}
    \item \textbf{Secret Key Rate:} The secret key $K$ satisfies $\frac{1}{n} H(K) \ge R - \epsilon$.
\end{enumerate}
The \emph{(strong) secret key capacity}, denoted by $C_{\mathrm{SK}}(X_\mathcal{M})$, is the supremum of all achievable strong SK rates. 
\end{definition}

Csisz\'{a}r and Narayan \cite{CN04} evaluated the secret key capacity $\csk(X_\cM)$ for a generic multiterminal source. More relevant to us is the alternate characterization of $\csk(X_\cM)$ given by Chan and Zheng in \cite{CZ10}. Let $\cP$ denote the set of all partitions of $\cM$ in at least two parts. The secret key capacity is then given by 
\begin{equation}
    \csk(X_\cM) = \min_{P\in\cP: |P| > 1} I_P(X_\cM),
    \label{eq:capacity-partition}
\end{equation}
where
\begin{equation}
    I_P(X_\cM) := \frac{1}{|P|-1} \left( \sum_{C \in P} H(X_C) - H(X_{\mathcal{M}}) \right).
    \label{eq:partition-info}
\end{equation}
Of relevance to us is the partition of $\cM$ into singletons, denoted by $\cS=\{\{1\},\{2\},\ldots,\{m\}\}$, which we shall henceforth refer to as the \emph{singleton partition}. Following the convention of \cite{mukherjee}, we shall refer to the sources where $\cS$ is a minimizer of the right-hand side of \eqref{eq:capacity-partition} as \emph{Type-$\cS$ sources}. Therefore, $\csk(X_\cM)=I_\cS(X_\cM)$ for Type-$\cS$ sources.

While $\csk(X_\cM)$ is defined under the asymptotic notion of strong secrecy (where error and leakage vanish as $n \to \infty$), there is a stricter non-asymptotic notion known as \emph{perfect secrecy}. A random variable $K$ is a perfect secret key if it satisfies exact recoverability ($\Pr(K_1 = \dots = K_m = K) = 1$), perfect secrecy ($I(K; F) = 0$), and perfect uniformity ($H(K) = \log |\mathcal{K}|$) for some blocklength $n$. It is immediate that a perfect secret key satisfies strong secrecy, but not vice versa. Accordingly, the perfect secret key capacity is upper-bounded by the usual strong secret key capacity.

We remark that henceforth, the phrase `secret key capacity' will be used to refer to the strong secret key capacity. Furthermore, we shall say that a perfect secret key achieves capacity if it has a rate equal to the secret key capacity. This work shows that, for instances of a special class of sources, called the hypergraphical sources, perfect secret keys achieve capacity.

\subsection{Hypergraphical source model}\label{sec:hyper}

A \emph{hypergraphical source} is defined using an underlying hypergraph $\mathcal{H} = (\mathcal{M}, \mathcal{E})$ whose vertex set is the set of parties $\mathcal{M}$. We allow $\cE$ to be a \emph{multiset}, i.e., $\cE$ is allowed to have repeated elements. Associated with each hyperedge $e \in \mathcal{E}$ are $n$ i.i.d. copies of a Bernoulli(1/2) random variable $\xi_e$. Furthermore, the collection of random variables $(\xi_e: e\in \cE)$ is mutually independent. The observation of party $i \in \mathcal{M}$ is the collection of variables associated with hyperedges incident to it, i.e.,  $X_i = (\xi_e : i \in e)$.
Consequently, the joint source is $X_{\mathcal{M}} = (\xi_e : e \in \mathcal{E})$. 

The combinatorial structure inherent in hypergraphical sources greatly simplifies the expression for the secret key capacity. This is especially evident in the case of the so-called \emph{$t$-uniform} hypergraphs. A hypergraph $\cH=(\cM,\cE)$ is said to be $t$-uniform if all hyperedges are of size $t$, i.e., $|e|=t$ for all $e\in\cE$. In particular, it is easy to see that the secret key capacity for Type-$\cS$ sources on $t$-uniform hypergraphs is given by 
\begin{equation}
    \csk(X_\cM)=I_\cS(X_\cM)=\frac{t-1}{m-1}|\cE|. \label{eq:typeS}
\end{equation}

A special case of the hypergraphical source is the \emph{pairwise independent network (PIN)} model \cite{pinorig,nitinrawat}, where the underlying hypergraph is a graph. Nitinawarat and Narayan \cite{nitinrawat} have shown that the PIN model admits a linear communication scheme which achieves a linear perfect secret key, via spanning tree packing of the underlying graph. In this work, we develop similar linear perfect secret key generation schemes for $t$-uniform hypergraphical sources by exploiting the combinatorial structure of the underlying hypergraph.

To proceed, we define some notation relevant to hypergraphical sources. Given a set $A\subseteq\cM$, we define the notation $\cE_A:=\{e\in \cE: e\cap A\neq\emptyset\}$. When $A=\{i\}$ (resp., $A=\{i,j\}$) we shall abuse notation and write $\cE_i$ instead of $\cE_{\{i\}}$ (resp., $\cE_{i,j}$ instead of $\cE_{\{i,j\}}$). For the $n$ random bits $\xi_e^n$ associated with the hyperedge $e$, we shall use the notation $\xi_{e(j)}$ to denote the $j$th i.i.d. copy of $\xi_e$. In fact, whenever necessary, we shall associate $n$ fictitious copies $e(1),e(2),\ldots,e(n)$ with every hyperegde $e\in\cE$, and associate $\xi_{e(j)}$ with $e(j)$. Furthermore, we define the set of all $n$ fictitious copies of the hyperedges $\cE^{(n)}=\{e(j):e\in\cE,1\leq j\leq n\}$. Next, given any $\cE'\subseteq\cE$, we define the tuples $\xi_{\cE'}:=(\xi_e:e\in\cE')$ and $\xi_{\cE'}^n:=(\xi_{e(j)}:e\in\cE', 1\leq j\leq n)$.

Hypergraphical sources in general admit linear communication schemes that consist of every party sending some linear combinations of the bits associated with the hyperedges incident on it. Formally, let $X_\cM$ be a hypergraphical source defined on the underlying hypergraph $\cH=(\cM,\cE)$. A public communication $F=(F_1,F_2,\ldots,F_m)$ is said to be \emph{linear} if for every $i\in\cM$, the party $i$ sends the communication $F_i=M_i\xi_\cE^n$,\footnote{All matrix operations and ranks considered in this paper are over $\mathbb{F}_2$.} where the matrix $M_i\in\{0,1\}^{r_i\times n|\cE|}$ for some integer $r_i\geq 0$. Furthermore, the matrices $M_i$ must satisfy $M_i(r,e(j))=1$ only if $e\in\cE_i$, to ensure that party $i$'s message involves linear combinations of its local observation only. Observe that the size of this linear communication is $\sum_{i\in\cM}r_i$ bits. A linear communication is said to achieve \emph{perfect omniscience} if every party $i$ can compute $\xi_\cE^n$ correctly with probability one using only $X_i^n$ and $F$. 

We next show that any perfect omniscience scheme can be leveraged to generate a perfect secret key. 

\begin{restatable}{lemma}{linkey}
    Let $X_{\cM}$ be a hypergraphical source defined on the underlying hypergraph $\cH=(\cM,\cE)$. For a blocklength $n\in\mathbb{N}$, let there exist a linear communication scheme of $r<n|\cE|$ bits achieving perfect omniscience. Then there exists a perfect secret key of rate at least $|\cE|-\frac{r}{n}$.
    \label{lem:master}
\end{restatable}

\begin{IEEEproof}
    See Appendix~\ref{app:hyper}.
\end{IEEEproof}

We conclude this section by showing how the notion of \emph{packing} can be used in tandem with the perfect secret key generation scheme of Lemma~\ref{lem:master} to increase the perfect secret key rate available from hypergraphical sources. A hypergraphical source $X_\cM$ on the underlying hypergraph $\cH=(\cM,\cE)$ is said to admit a packing for blocklength $n$, if there exists disjoint subsets $\cE_{(1)},\cE_{(2)},\ldots,\cE_{(s)}\subseteq\cE^{(n)}$. Then, $\cH_{(l)}:=(\cM,\cE_{(l)})$ can be viewed as a hypergraph in its own right. Moreover, we define $X_{i_{(l)}}:=(\xi_{e(j)}: i\in e, e(j)\in\cE_{(l)})$, and consequently $X_{\cM_{(l)}}:=(X_{i_{(l)}}:i\in\cM)$ is a hypergraphical source. 

\begin{restatable}{lemma}{packedkeys}
    Let $X_\cM$ be a hypergraphical source on the underlying hypergraph $\cH=(\cM,\cE)$. For a blocklength $n$, let $\cH$ admit a packing, such that the hypergraphical source $X_{\cM_{(l)}}, 1\leq l\leq s$, can generate a perfect secret key of $k_l$ bits using a blocklength of one. Then, $X_\cM$ can generate a perfect secret key of rate $\frac{\sum_{l=1}^sk_l}{n}$.
    \label{lem:packedkeys}
\end{restatable}

\begin{IEEEproof}
    See Appendix~\ref{app:hyper}.   
\end{IEEEproof}

\section{Perfect secret keys achieving capacity for complete $t$-uniform hypergraphs}\label{sec:tunif}

This section considers hypergraphical sources on the complete $t$-uniform hypergraph $K_{m,t}$. The complete $t$-uniform hypergraph possesses every possible hyperedge of size $t$, i.e., $\mathcal{E}=\{e\subset\cM:|e|=t\}$. $K_{m,t}$ is known to be Type-$\cS$ \cite{mukherjee}, and this leads to a simple expression for its capacity.

\begin{lemma}
    The secret key capacity of the hypergraphical source on $K_{m,t}$ is given by $\frac{t-1}{m-1}\binom{m}{t}$.
    \label{lem:typescomp}
\end{lemma}

\begin{IEEEproof}
    This follows immediately from the fact that $K_{m,t}$ is Type-$\cS$ \cite[Corollary~19]{mukherjee}, the capacity expression for Type-$\cS$ hypergraphical sources in \eqref{eq:typeS}, and noting $|\cE|=\binom{m}{t}$.
    
\end{IEEEproof}

To obtain the perfect secret key achieving capacity, we shall first obtain a packing of $K_{m,t}$ into what we refer to as \emph{star hypergraphs}. 

\begin{definition}[Star Hypergraphs $S_{\cH,i}$]
Consider a hypergraphical source $X_\cM$ defined on a hypergraph $\cH=(\cM,\cE)$, and fix $i\in\cM$. Then, the star hypergraph \emph{anchored} at $i$ is the subhypergraph $S_{\cH,i}=(\cM,\cE_i)$. Furthermore, the hypergraphical source associated with $S_{\cH,i}$ is denoted by $X_{\cM_{S_i}}:=(X_{1_{S_i}},\ldots,X_{m_{S_i}})$ where $X_{j_{S_i}}:=(\xi_e:i,j\in e)$, for all $j\in\cM$.  
\end{definition}
For the remainder of this section, we shall use $S_i$ to denote $S_{K_{m,t},i}$ for notational convenience.  

Our perfect secret key generation scheme for $K_{m,t}$ proceeds in two phases. We first show that for a blocklength $n=t$, the hypergraphical source on $K_{m,t}$ admits a packing into star hypergraphs. Then, we construct a perfect secret key generation scheme for each star hypergraph. Combining these results with Lemma~\ref{lem:packedkeys} we are able to show that our key generation scheme gives perfect secret keys achieving capacity.

\begin{restatable}{lemma}{sourcedecomp}
\label{lemma:decomposition}
The hypergraphical source on $K_{m,t}$, for a blocklength $t$, admits a packing $\cE_{(1)},\cE_{(2)},\ldots,\cE_{(m)}\subseteq\cE^t$, where $K_{{m,t}_{(l)}}=(\cM,\cE_{(l)}), 1\leq l\leq m,$ is the star hypergraph $S_l$.
\end{restatable}

\begin{IEEEproof}
    See Appendix~\ref{app:tunif}.
\end{IEEEproof}

Next, we construct a linear communication achieving perfect omniscience for the source $X_{\cM_{S_i}}$ on $S_i$.

\begin{lemma}
    For the source $X_{\cM_{S_i}}$ on $S_i$, there exists a linear communication of $\binom{m-1}{t-1}-\binom{m-2}{t-2}$ bits achieving perfect omniscience. Hence, $S_i$ can generate a perfect secret key of rate $\binom{m-2}{t-2}$.
    \label{lem:potstar}
\end{lemma}

\begin{IEEEproof}
    We construct a communication scheme where only party $i$ sends messages. Fix any $j\neq i$. For any $e\in\cE_i\setminus\cE_{i,j}$, the party $i$ sends the message $(\oplus_{k\in e\setminus\{i\}}\xi_{(e\cup\{j\})\setminus\{k\}})\oplus\xi_e$. The communication scheme is linear by construction. Furthermore, the number of bits sent is $|\cE_i\setminus\cE_{i,j}|=\binom{m-1}{t-1}-\binom{m-2}{t-2}$. We first show that this communication scheme achieves perfect omniscience by showing that every $k\in\cM$ recovers $\xi_{\cE_i}$.

    Obviously $i$ already knows $\xi_{\cE_i}$. Next, note that for any $e\in\cE_i\setminus\cE_{i,j}$, the party $j$ can recover $\xi_e$ from the message $(\oplus_{k\in e\setminus\{i\}}\xi_{(e\cup\{j\})\setminus\{k\}})\oplus\xi_e$. Thus, $j$ perfectly recovers $\xi_{\cE_i}$. Next, fix any $k\in\cM\setminus\{i,j\}$. Since $j$ can recover $\xi_{\cE_i}$ using the communication and $\xi_{\cE_{i,j}}$, it is enough to show that $k$ can recover $\xi_{\cE_{i,j}}$. Observe that $k$ already has access to $\xi_e$ for any $e\in\cE_{i,j}$ such that $k\in e$. Now, fix any $e\in\cE_{i,j}$ such that $k\notin e$. Consider the hyperedge $e'=(e\cup\{k\})\setminus\{j\}\in\cE_i\setminus\cE_{i,j}$. Observe that $k$ has access to $\xi_{(e'\cup\{j\})\setminus\{l\})}$ for every $l\in e'\setminus\{i,k\}$. Then, $k$ can recover $\xi_e$ from the communication message $(\oplus_{l\in e'\setminus\{i\}}\xi_{(e\cup\{j\})\setminus\{l\}})\oplus\xi_{e'}$, as every hyperedge involved in this message contains $k$ except $e$. 

    Following perfect omniscience, the key generation scheme in Lemma~\ref{lem:master} gives a perfect secret key of rate $|\cE_i|-(\binom{m-1}{t-1}-\binom{m-2}{t-2})=\binom{m-2}{t-2}$.\footnote{In fact, for $S_i$ we can explicitly define the perfect key to be $(\xi_e:e\in\cE_{i,j})$. This key is obviously perfectly recoverable and uniform. The argument that it is perfectly secured from the communication follows by using a similar rank-based argument used in the proofs of Lemmas~\ref{lem:rank} and \ref{lem:master}.}
\end{IEEEproof}

We digress here to note that the perfect secret keys generated in Lemma~\ref{lem:potstar} achieve capacity for $X_{\cM_{S_i}}$. Using \eqref{eq:typeS}, we see that 
\begin{gather*}
    I_\cS(X_{\cM_{S_i}})=\frac{t-1}{m-1}|\cE_i|=\frac{t-1}{m-1}\binom{m-1}{t-1}=\binom{m-2}{t-2}.
\end{gather*}
An application of Lemma~\ref{lem:typescomp} immediately proves that $S_i$ is Type-$\cS$ and the perfect secret key thus generated achieves capacity.

The packing scheme of Lemma~\ref{lemma:decomposition} can now be combined with the key generation scheme for $S_i$ in Lemma~\ref{lem:potstar} to generate perfect secret keys for $K_{m,t}$ achieving capacity. 

\begin{theorem}\label{thm:kmt}
There is an explicit perfect secret key generation scheme for $K_{m,t}$ that achieves capacity.
\end{theorem}

\begin{IEEEproof}
Lemma~\ref{lemma:decomposition} gives a packing $\cE_{(1)},\ldots,\cE_{(m)}\subseteq\cE^n$ of $K_{m,t}$ for a blocklength $n=t$. Furthermore, $K_{{m,t}_{(i)}}=(\cM,\cE_{(i)})$ is exactly the hypergraph $S_i$, for every $i\in\cM$. Each of them can thus generate a perfect secret key of $\binom{m-2}{t-2}$ bits using the scheme of Lemma~\ref{lem:potstar}. Finally, by Lemma~\ref{lem:packedkeys}, this generates a perfect secret key of rate
\begin{gather}
    \frac{m}{t}\binom{m-2}{t-2}=\frac{t-1}{m-1}\frac{m(m-1)}{t(t-1)}\binom{m-2}{t-2}=\frac{t-1}{m-1}\binom{m}{t},
\end{gather}
which by Lemma~\ref{lem:typescomp} equals the secret key capacity of $K_{m,t}$.
\end{IEEEproof}

\section{Secret key generation scheme for $3-$uniform hypergraphs}\label{sec:threeunif}

In this section, we provide a new perfect secret key generation scheme for  $3$-uniform hypergraphs (not necessarily complete), and we identify certain classes of 3-uniform hypergraphs where our scheme achieves capacity.

We begin by defining the notion of \emph{projection} of a star hypergraph.
\begin{definition}
    Let $\cH=(\cM,\cE)$ be a hypergraph, and let $i\in\cM$. The projection of the star hypergraph $S_{\cH,i}$, denoted by $P_{\cH,i}$ is the hypergraph with vertex set $\cM\setminus\{i\}$ and set of hyperedges $\{e\setminus\{i\}:e\in\cE_i\}$.
    \label{def:project}
\end{definition}
The main ingredient in our construction is a type of $3$-uniform hypergraph which we call \emph{cycle-inducing} hypergraph. 
\begin{definition}
    A $3$-uniform hypergraph $\cH=(\cM,\cE)$ is said to be cycle-inducing if there exists $i\in\cM$ such that $\cH=S_{\cH,i}$ (i.e., $\cE=\cE_i$), and $P_{\cH,i}$ is a cycle\footnote{For definitions of standard graph theory terms such as a cycle, refer to any standard textbook on graph theory such as \cite{diestel}.} on the vertex set $\cM\setminus\{i\}$.

    We call $i\in\cM$ as the \emph{anchor}. 
    \label{def:cycind}
\end{definition}

The key result leading to our construction is the following proposition.

\begin{restatable}{proposition}{prop}\label{lem:induced_cycle}
    Let $\cH=(\cM,\cE)$ be a cycle-inducing $3$-uniform hypergraph with the anchor $i$. There exists a linear communication scheme of $m-3$ bits achieving perfect omniscience. As a consequence, one can generate a perfect secret key of 2 bits. 
\end{restatable}

\begin{IEEEproof}
    See Appendix~\ref{app:3unif}.
\end{IEEEproof}

Before describing our generic scheme for $3$-uniform hypergraphs, we require one standard notion from graph theory, stated below.
\begin{definition}
    A (multi)graph\footnote{Multigraph simply means that we allow the graph to have repeated edges.} $G$ admits a \emph{Hamiltonian packing} of size $r$ if there exists at least $r$ edge disjoint Hamiltonian cycles\footnote{Recall that a cycle subgraph of a graph is called Hamiltonian if it includes all the vertices of the graph -- see for example Chapter~10 of \cite{diestel}.} $C^{(1)},\ldots, C^{(r)}$ which are subgraphs of $G$.

    If the set of cycles $C^{(1)},\ldots, C^{(r)}$ exhausts all the edges of the graph, then we say that $C^{(1)},\ldots, C^{(r)}$ is a \emph{Hamiltonian decomposition}. 
    \label{def:Hamilton}
\end{definition}

The following theorem now describes the perfect secret key generation scheme for $3$-uniform hypergraphs.

\begin{restatable}{theorem}{threeunifscheme}
    Let $\cX_\cM$ be a source on a 3-uniform hypergraph $\mathcal{H} = (\mathcal{M}, \mathcal{E})$. Let $A\subseteq\cM$ be such that for every $i\in A$, there exists a positive integer $k_i$ such that $k_i$ copies of $P_{\cH,i}$ admit a Hamiltonian packing of size $p_i$. For each edge $e\in\cE$ define the edge reuse number $t_e = \sum_{i\in A} k_i \mathbf{1}\{i \in e\}$.\footnote{$\mathbf{1}\{\cdot\}$ denotes the indicator function.} Then, using a blocklength $n=\max_{e\in \mathcal{E}} t_e$, perfect secret keys of rate $\frac{2}{n}\sum_{i\in A}p_i$ can be generated.
    \label{th:3unif}
\end{restatable}

\begin{IEEEproof}
    See Appendix~\ref{app:3unif}.
\end{IEEEproof}

We now identify three different classes of $3$-uniform hypergraphs for which the scheme described in Theorem~\ref{th:3unif} is capacity achieving.

First, we consider the class of complete $3$-uniform hypergraph on $m$ vertices. 
\begin{restatable}{corollary}{typea}\label{lem:generic_complete}
    For the complete $3$-uniform hypergraph $K_{m ,3}$ on $m$ vertices, the scheme of Theorem~\ref{th:3unif} with the set $A = \cM$ achieves capacity. 
\end{restatable}

\begin{IEEEproof}
    See Appendix~\ref{app:3unif}.
\end{IEEEproof}

Next, we consider a family of $3$-uniform hypergraphs $\cH=(\cM,\cE)$ for which there exists $i\in\cM$ such that $\cE_i=\cE$ and the graph $P_{\cH,i}$ admits a Hamiltonian decomposition. An interesting class of graphs admitting a Hamiltonian decomposition are Paley graphs \cite{ALSPACH2012113}. More such classes can be found in \cite{ALSPACH2012113}.

\begin{restatable}{corollary}{typeb}\label{lem:generic_hamdecomp}
    Consider a hypergraphical source on a $3$-uniform $\mathcal{H} = (\mathcal{M}, \mathcal{E})$ for which there exists a vertex $i \in \mathcal{M}$ satisfying $\cE_i=\cE$ and the graph $P_{\cH,i}$ admits a Hamiltonian decomposition. Then, the perfect secret key generation scheme of Theorem~\ref{th:3unif} is capacity achieving. Furthermore, such a hypergraphical source is Type-$\cS$. 
\end{restatable}

\begin{IEEEproof}
    See Appendix~\ref{app:3unif}.
\end{IEEEproof}

We conclude our set of examples of $3$-uniform hypergraphs where the scheme of Theorem~\ref{th:3unif} achieves capacity with a class of hypergraphs we term \textit{Hollow 3D Kite hypergraphs}. A Hollow 3D Kite hypergraph $\cH=(\cM,\cE)$ is defined as follows. The number of vertices $m=4r+1$ for some $r\geq 1$, and let $V_1 = \{1, \dots, 2r+1\}$ and $V_2 = \{2r+2, \dots, 4r+1\}$. The set of hyperedges is then defined as $\mathcal{E} = \left\{ \{i, j, k\} : \{i, j\} \subseteq V_1, k \in V_2 \right\}$. 

\begin{restatable}{corollary}{typec}\label{lem:generic_kite}
    Consider a hypergraphical source on a $3$-uniform hypergraph $\cH=(\cM,\cE)$ which is a Hollow 3D Kite on $m=4r+1$ vertices. Then, the scheme of Theorem~\ref{th:3unif} gives perfect secret keys achieving capacity. Furthermore, the source is Type-$\cS$.
\end{restatable}

\begin{IEEEproof}
    See Appendix~\ref{app:3unif}.
\end{IEEEproof}

\section{Conclusion}\label{sec:conc}

In this work, we gave explicit capacity achieving perfect secret key generation schemes for the complete $t$-uniform hypergraph and certain classes of $3$-uniform hypergraphs. Our schemes are an attempt to extend the spanning tree packing scheme of Nitinawarat and Narayan \cite{nitinrawat} meant for PIN models on graphs. Unfortunately, there is no unique generalization of spanning trees to the setting of hypergraphs. Hence, for our scheme for $K_{m,t}$, we pack the hypergraph with star hypergraphs, which serve as the proxy for the spanning trees, and generate perfect secret keys for each star hypergraph. Similarly, the scheme for $3$-uniform hypergraphs uses cycle inducing $3$-uniform hypergraphs as stand-ins for spanning trees. We believe that a good generalization of spanning trees will shed further light on the problem of explicit perfect secret key generation schemes for hypergraphical sources. 

In this regard, we believe that the so-called \emph{minimally topologically connected} hypergraphs studied in \cite{HTKZPW} are a good candidate for generalizations of spanning trees. Indeed, the common randomness generation scheme for minimally topologically connected hypergraphs studied in \cite{HTKZPW} coupled with Lemma~\ref{lem:master} gives an explicit construction of perfect secret keys achieving capacity. We conjecture that every topologically connected hypergraph can be suitably packed by minimally topologically connected hypergraphs such that invoking Lemma~\ref{lem:packedkeys} will give a perfect secret key achieving capacity. 

Furthermore, note that we have a scheme for $3$-uniform hypergraphs which achieves capacity for non-trivial classes of hypergraphs other than the complete hypergraph. A parallel direction of research would be to extend our scheme to $t$-uniform hypergraphs. This would require a generalization of the cycle to hypergraphs, and we conjecture that sphere triangulations is the correct generalization to look at in this case. 

\begin{appendices}
\section{Omitted proofs of Section~\ref{sec:hyper}}\label{app:hyper}

To prove Lemma~\ref{lem:master}, we first need the following technical lemma.

\begin{lemma}
    Let $s\in\mathbb{N}$ and let $Y=(Y_1,Y_2,\ldots,Y_s)^T\sim\unif\{\{0,1\}^s\}$. Let $M\in\{0,1\}^{l\times s}$ be a matrix whose rows we denote as $M_1,M_2,\ldots,M_l$. For any $1\leq l'<l$ if $\spanv\{M_1,\ldots,M_{l'}\}\cap\spanv\{M_{l'+1},\ldots,M_l\}=\{0\}$,\footnote{$0$ here denotes the vector $(\underbrace{0,0,\ldots,0}_{s\text{ times}})$.} then $I(M_1Y,\ldots,M_{l'}Y;M_{l'+1}Y,\ldots,M_lY)=0$. Furthermore, $H(M_{l'+1}Y,\ldots,M_lY)=\dim(\spanv\{M_{l'+1},\ldots,M_l\})$. 
    \label{lem:rank}
\end{lemma}

\begin{IEEEproof}
    This result follows by using the standard result that $H(AY)=\rank(A)$ given any matrix $A\in\{0,1\}^{r\times s}$ -- see, for example, Theorem~7.3 of \cite{yeung2006}. Using the above, consider matrices $A$ (resp. $B$) whose rows are $M_1,\ldots,M_{l'}$ (resp. $M_{l'+1},\ldots,M_l$). Firstly, observe that the condition $\spanv\{M_1,\ldots,M_{l'}\}\cap\spanv\{M_{l'+1},\ldots,M_l\}=\{0\}$ implies that $\rank(M)=\rank(A)+\rank(B)$. 

    Then, by the result stated above, we have 
    $$H(M_{l'+1}Y,\ldots,M_lY)=H(BY)=\rank(B)=\dim(\spanv\{M_{l'+1},\ldots,M_l\}).$$ Also, 
    \begin{align*}
        I(M_1Y,\ldots,M_{l'}Y;M_{l'+1}Y,\ldots,M_lY) & = I(AY;BY)\\
                                                     & = H(AY,BY)-H(AY)-H(BY)\\
                                                     & = H(MY)-H(AY)-H(BY)\\
                                                     & = \rank(M)-\rank(A)-\rank(B)\\
                                                     & = 0.
    \end{align*}
\end{IEEEproof}

Lemma~\ref{lem:rank} is now leveraged to prove Lemma~\ref{lem:master}. 

\linkey*
\begin{IEEEproof}
    Since the communication is linear and of size $r$ bits, we can write $F=M\xi_\cE^n$, where $M\in\{0,1\}^{r\times n|\cE|}$. Hence, $\rank(M)\leq r$ as $r<n|\cE|$. Let $M_1,M_2,\ldots,M_r$ be the rows of $M$, and hence $\dim(\spanv\{M_1,\ldots,M_r\})=r'\leq r$. Let $M'_1,M'_2,\ldots,M'_{r'}$ be a basis of $\spanv\{M_1,\ldots,M_r\}$. Then, this can be extended to a basis $\{M'_1,M'_2,\ldots,$ $M'_{n|\cE|}\}$ of $\{0,1\}^{n|\cE|}$. We claim that $K=(M'_j\xi_\cE^n:r'+1\leq j\leq n|\cE|)$ is the necessary perfect secret key. The key alphabet $\cK$ is thus $\{0,1\}^{n|\cE|-r'}$.

    To see that the claim holds, firstly observe that $K$ is perfectly recoverable as every party $i\in\cM$ has access to $\xi_\cE^n$ thanks to perfect omniscience. Lemma~\ref{lem:rank} shows that $I(K;F)=0$. Furthermore, as $\{M'_{r'+1},\ldots,M'_{n|\cE|}\}$ is linearly independent, Lemma~\ref{lem:rank} shows that $H(K)=n|\cE|-r'=\log|\cK|$. Thus, the key $K$ is perfectly uniform, and hence $K$ is a perfect secret key. The rate of $K$ is thus $\frac{1}{n}H(K)=|\cE|-\frac{r'}{n}\geq |\cE|-\frac{r}{n}$.
\end{IEEEproof}

\packedkeys*
\begin{IEEEproof}
    Let $K_l, 1\leq l\leq s$, be the perfect secret key of $k_l$ bits generated by $X_{\cM_{(l)}}$ through public communication $F_{(l)}$. Then, $H(K_l)=k_l$ and $I(K_l;F_{(l)})=0$. Denote the overall public communication by $F=(F_{(l)}:1\leq l\leq s)$ and define $K=(K_l:1\leq l\leq s)$. We claim that $K$ is the required perfect secret key. 
    
    To prove the claim, firstly observe that $K$ is perfectly recoverable, since the individual $K_l$s are perfectly recoverable. Next, note that $\xi_{\cE_{(1)}},\xi_{\cE_{(2)}},\ldots,\xi_{\cE_{(s)}}$ are mutually independent as $\cE_{(1)},\cE_{(2)},\ldots,\cE_{(l)}$ are disjoint sets. As a consequence, $K_1,\ldots,K_s$ are mutually independent, and so are $F_{(1)},F_{(2)},\ldots,F_{(l)}$. Hence, $H(K)=\sum_{l=1}^sH(K_l)=\sum_{l=1}^sk_l$, and so $K$ is uniformly distributed over the key alphabet $\{0,1\}^{\sum_{l=1}^sk_l}$. Finally, the mutual independece of $K_l$s and $F_{(l)}$s coupled with the fact that $I(K_l;F_{(l)})=0$ for all $1\leq l\leq s$ ensures that $I(K;F)=0$, and hence $K$ is perfectly secure. The rate of $K$ is thus $\frac{1}{n}H(K)=\frac{1}{n}\sum_{l=1}^sk_l$ as required.
\end{IEEEproof}

\section{Omitted proofs of Section~\ref{sec:tunif}}\label{app:tunif}
\sourcedecomp*
\begin{IEEEproof}
We obtain the packing as follows. Fix any $e\in\cE$ and order its vertices in the usual ascending order. In other words, let $e=\{v_1,v_2,\ldots,v_t\}$ where $v_1<v_2<\ldots v_t$. Then assign $e(l)$ to $\cE_{(v_l)}$. For example, if $m=4,t=3,$ and $e=\{1,2,4\}$, then $e(1)$ is assigned to $\cE_{(1)}$, $e(2)$ is assigned to $\cE_{(2)}$, and $e(3)$ is assigned to $\cE_{(4)}$. Observe that by construction $\cE_{(1)},\cE_{(2)},\ldots,\cE_{(m)}$ is indeed a packing as any $e(j), e\in\cE, 1\leq j\leq t$ is assigned to exactly one $\cE_{(l)}$. It remains to show that for any $1\leq l\leq m$, the set $\cE_{(l)}$ consists of exactly one fictitious copy of every edge in $\cE_l$. 

To proceed, fix any $l\in\cM$ and $e\in\cE_l$. It is enough to show that for exactly one $1\leq j\leq t$, $e(j)\in\cE_{(l)}$. Now, let $e=\{v_1,v_2,\ldots,v_t\}$. Then, as $l\in e$, there exists exactly one $1\leq j\leq t$ such that $v_j=l$. Thus, by construction, only $e(j)$ gets assigned to $\cE_{(l)}$ as required. 
\end{IEEEproof}

\section{Omitted Proofs of Section~\ref{sec:threeunif}}\label{app:3unif}
\prop*
\begin{IEEEproof}
 For $m=4$, we simply have the anchor send the XOR of all the bits associated with all three hyperedges. Then, every party, having access to at least two hyperedges, recovers the random bits associated with all the hyperedges. Hence, we turn our attention to the case where $m = k+1,  k\geq 4$. 
 
 We only need to show the necessary $m-3$ bit linear communication achieving perfect omniscience. The 2-bit perfect secret key then follows immediately from Lemma~\ref{lem:master} by noting $|\cE|=m-1$ as $\cH$ is cycle inducing. To define the linear communication, without loss of generality, let party $k+1$ be the anchor, and let the set of edges of $P_{\cH,k+1}$ be $\{\{j,j+1\}:1\leq j\leq k-1\}\cup\{\{1,k\}\}$. We introduce the following notation for the hyperedges in $\cE$: for each $1\leq j\leq k-1, e_j := \{k+1, j, j+1\}$ and $e_k = \{k+1, k, 1\}$. 

\textbf{Case I}: $k=3l$ or $k=3l+2$

The protocol is as follows: the anchor $k+1$ sends the following $k-2$ messages: $\xi_{e_1}\oplus \xi_{e_2} \oplus \xi_{e_3}$, $\xi_{e_2}\oplus \xi_{e_3}\oplus \xi_{e_4},\ldots,\xi_{e_{k-2}}\oplus \xi_{e_{k-1}} \oplus \xi_{e_k}$. The number of bits of messages thus sent is $k-2=m-3$.

To show perfect omniscience, first, consider some party $j\neq 1$. By virtue of being a member of those hyperedges, party $j$ has access to the bits $\xi_{e_{j-1}}, \xi_{e_j}$. From these, they can recover the bits $\xi_{e_{j-2}}, \xi_{e_{j+1}}$ from the communications $\xi_{e_{j-2}}\oplus \xi_{e_{j-1}} \oplus \xi_{e_j}$ and $\xi_{e_{j-1}}\oplus \xi_{e_{j}} \oplus \xi_{e_{j+1}}$, respectively. Inductively, they can now recover all the other bits. Now, for party $1$, they have access to edges $\xi_{e_1}$ and $\xi_{e_k}$. By adding the communications $\xi_{e_1}\oplus \xi_{e_2} \oplus \xi_{e_3}$ and $\xi_{e_2}\oplus \xi_{e_3} \oplus \xi_{e_4}$ along with the knowledge of $\xi_{e_1}$, they can recover $\xi_{e_4}$. Inductively, they can recover every bit of the form $\xi_{e_{3c+1}}$. 

If $k=3l$, the last bit party $1$ recovers in this sequence is $\xi_{e_{3l-2}}$, which allows them to recover $\xi_{e_{3l-1}}$ from the communication $\xi_{e_{3l-2}}\oplus \xi_{e_{3l-1}} \oplus \xi_{e_{3l}}$. Party $1$ now has the same information as party $3l$, which we have shown to be able to recover everything. Therefore, party $1$ can recover all the information.

Similarly, if $k=3l+2$, the last bit recovered in this sequence is $\xi_{e_{3l+1}}$, which allows party $1$ to recover all the knowledge of party $3l+2$, which suffices again by the same argument as above. 

\textbf{Case II:} $k=3l+1$

The protocol is as follows: the anchor $k+1$ sends the messages $\xi_{e_1}\oplus \xi_{e_2} \oplus \xi_{e_3},\ldots,\xi_{e_{k-3}}\oplus \xi_{e_{k-2}} \oplus \xi_{e_{k-1}}$, and finally the message $\xi_{e_{k-1}}\oplus \xi_{e_k} \oplus \xi_{e_1}$. The number of bits of messages thus sent is $k-2=m-3$. 
    
To show recovery, first, consider some party $j \notin \{1, k\}$. Party $j$ has access to $\xi_{e_{j-1}}, \xi_{e_j}$. Using these and the communication, an argument exactly same as that in the previous case shows that they can recover $\xi_{e_1}, \ldots,\xi_{e_{k-1}}$. Now, possessing both $\xi_{e_1}$ and $\xi_{e_{k-1}}$, they can use the last message $\xi_{e_{k-1}} \oplus \xi_{e_k} \oplus \xi_{e_1}$ to recover $\xi_{e_k}$. Thus, they recover all bits. 
    
Party $1$, having access to $\xi_{e_1}, \xi_{e_k}$, immediately recovers $\xi_{e_{k-1}}$ from the last message $\xi_{e_{k-1}}\oplus \xi_{e_k} \oplus \xi_{e_1}$. As in Case I, party $1$ also recovers bits associated with hyperedges of the form $\xi_{e_{3c+1}}$. Since $k=3l+1$, and so $k-3=3(l-1)+1$, the last bit party $1$ recovers in this way is $\xi_{e_{k-3}}$. Now, using $\xi_{e_{k-3}}$ and $\xi_{e_{k-1}}$, party $1$ uses the message $\xi_{e_{k-3}}\oplus \xi_{e_{k-2}} \oplus \xi_{e_{k-1}}$ to recover $\xi_{e_{k-2}}$. Party $1$ thus recovers the view of party $k-1$, which has already been shown to recover all the bits.  
    
For party $k$, we show that it can recover the view of party $1$, which has already been shown to recover everything. Using the last message $\xi_{e_{k-1}}\oplus \xi_{e_k} \oplus \xi_{e_1}$, they recover $\xi_{e_1}$. Together with $\xi_{e_k}$, this recovers the view of party $1$ as required.
\end{IEEEproof}

\threeunifscheme*

\begin{IEEEproof}
    Set $\delta:=\sum_{i\in A}p_i$. We begin with the claim that for blocklength $n$ given in the hypothesis of Theorem~\ref{th:3unif}, there exists a packing of $\cH$ into $\cE_{(1)},\ldots,\cE_{(\delta)}\subseteq\cE^n$ such that each of the hypergraphs $\cH_{(l)}, 1\leq l\leq\delta,$ are cycle inducing. Therefore, by Proposition~\ref{lem:induced_cycle}, each $\cH_{(l)}$ generates 2 bits of perfect secret key. This leads to a perfect secret key of rate $\frac{2}{n}\sum_{i\in A}p_i$ by using the Lemma~\ref{lem:packedkeys}.

    To complete the proof, it remains to prove the claim that the required packing exists. To facilitate the proof, we first order the set $A$ as $A=\{u_1,u_2,\ldots,u_{|A|}\}$, with $u_1<u_2<\ldots<u_{|A|}$. Next, we reindex the sets $\cE_{(l)}$ as $\cE_{(i,j_i)}, 1\leq i\leq |A|, 1\leq j_i\leq p_{u_i}$. Fix any $e\in\cE$, and define $A_e=\{i\in A:i\in e\}$, and let $A_e=\{u_{l_1},u_{l_2},\ldots,u_{l_{|A_e|}}\}$, where $l_1<l_2<\ldots<l_{|A_e|}$. Next, for any $l\in A$, let $C_l^{(1)},\ldots,C_l^{(p_l)}$ denote the $p_l$ cycles obtained by the Hamiltonian packing of $k_l$ copies of $P_{\cH,l}$. We say that $e$ takes part in the cycle $C_l^{s}$ if $e=\{l\}\cup e'$, where $e'$ is some edge of the cycle $C_l^{(s)}$. 

    We build the sets $\cE_{(i,j_i)}$ as follows. For any $1\leq q\leq|A_e|,$ and any $1\leq j_{l_q}\leq p_{u_{l_q}}$, assign the fictitious copy $e(\sum_{r=1}^{q-1}k_{l_r}+j_{l_q})$ to $\cE_{(l_q,j_{l_q})}$ provided $e$ takes part in $C_{u_{l_q}}^{(j_{l_q})}$. Then, by construction, the sets $\cE_{(i,j_i)}$ are disjoint. Furthermore, we claim that $e$ never runs out of fictitious copies needed for the assignment process described above. To see this, note that for any $l_q, 1\leq q\leq |A_e|$, the hyperedge $e$ takes part in at most $k_{u_{l_q}}$ of the cycles $C_{u_{l_q}}^{(s)}, 1\leq s\leq p_{u_{l_q}}$. Thus, the maximum number of copies of $e$ required by the assignment process is $\sum_{q=1}^{|A_e|}k_{l_q}=\sum_{l\in A}k_l\mathbf{1}\{l\in e\}=t_e\leq n$. Hence, $e$ never runs out of fictitious copies. Finally, observe that the assignment process ensures that $\cE_{(i,j_i)}$ gets exactly one fictitious copy of every $e$ taking part in $C_{u_i}^{(j_i)}, 1\leq j_i\leq p_{u_i}, 1\leq i\leq|A|$. Hence, with a slight abuse of notation, $\cH_{(i,j_i)}$ is cycle inducing with anchor at $u_i$, as required. This completes the proof of the claim, and hence, the proof of Theorem~\ref{th:3unif}.
\end{IEEEproof}

\begin{figure}[htbp]
\centering

\begin{subfigure}{0.32\textwidth}
\centering
\begin{tikzpicture}[scale=1.0, every node/.style={circle,draw,inner sep=2pt}]
    \node (1) at (90:2) {1}; \node (2) at (0:2) {2};
    \node (3) at (-90:2) {3}; \node (4) at (180:2) {4};
    \draw[copyone] (1)--(2)--(3)--(4)--(1);
\end{tikzpicture}
\caption{Cycle 1}
\end{subfigure}
\hfill
\begin{subfigure}{0.32\textwidth}
\centering
\begin{tikzpicture}[scale=1.0, every node/.style={circle,draw,inner sep=2pt}]
    \node (1) at (90:2) {1}; \node (2) at (0:2) {2};
    \node (3) at (-90:2) {3}; \node (4) at (180:2) {4};
    \draw[copyone] (1)--(3); \draw[copyone] (2)--(4);
    \draw[copytwo] (4)--(1); \draw[copytwo] (3)--(2);
\end{tikzpicture}
\caption{Cycle 2}
\end{subfigure}
\hfill 
\begin{subfigure}{0.32\textwidth}
\centering
\begin{tikzpicture}[scale=1.0, every node/.style={circle,draw,inner sep=2pt}]
    \node (1) at (90:2) {1}; \node (2) at (0:2) {2};
    \node (3) at (-90:2) {3}; \node (4) at (180:2) {4};
    \draw[copytwo] (1)--(4)--(2)--(3)--(1);
\end{tikzpicture}
\caption{Cycle 3}
\end{subfigure}

\caption{Hamiltonian cycle packing for $K_4$. The edges originating from the two distinct copies are respectively coloured red and blue.}
\label{fig:K4-final}

\vspace{1cm} 

\begin{subfigure}{0.32\textwidth}
\centering
\begin{tikzpicture}[scale=0.85, every node/.style={circle,draw,inner sep=2pt}]
    \node (1) at (90:2) {1}; \node (2) at (30:2) {2};
    \node (3) at (-30:2) {3}; \node (4) at (-90:2) {4};
    \node (5) at (-150:2) {5}; \node (6) at (150:2) {6};
    \draw[copyone] (1)--(4)--(3)--(5)--(2)--(6)--(1);
\end{tikzpicture}
\caption{Cycle 1}
\end{subfigure}
\hfill
\begin{subfigure}{0.32\textwidth}
\centering
\begin{tikzpicture}[scale=0.85, every node/.style={circle,draw,inner sep=2pt}]
    \node (1) at (90:2) {1}; \node (2) at (30:2) {2};
    \node (3) at (-30:2) {3}; \node (4) at (-90:2) {4};
    \node (5) at (-150:2) {5}; \node (6) at (150:2) {6};
    \draw[copyone] (1)--(5)--(6)--(4)--(2)--(3)--(1);
\end{tikzpicture}
\caption{Cycle 2}
\end{subfigure}
\hfill
\begin{subfigure}{0.48\textwidth}
\centering
\begin{tikzpicture}[scale=0.85, every node/.style={circle,draw,inner sep=2pt}]
    \node (1) at (90:2) {1}; \node (2) at (30:2) {2};
    \node (3) at (-30:2) {3}; \node (4) at (-90:2) {4};
    \node (5) at (-150:2) {5}; \node (6) at (150:2) {6};
    \draw[copyone] (1)--(2); \draw[copyone] (3)--(6);
    \draw[copytwo] (2)--(4)--(3); \draw[copytwo] (6)--(5)--(1);
\end{tikzpicture}
\caption{Cycle 3}
\end{subfigure}
\hfill
\begin{subfigure}{0.48\textwidth}
\centering
\begin{tikzpicture}[scale=0.85, every node/.style={circle,draw,inner sep=2pt}]
    \node (1) at (90:2) {1}; \node (2) at (30:2) {2};
    \node (3) at (-30:2) {3}; \node (4) at (-90:2) {4};
    \node (5) at (-150:2) {5}; \node (6) at (150:2) {6};
    \draw[copytwo] (1)--(6)--(4); \draw[copyone] (4)--(5);
    \draw[copytwo] (5)--(2)--(3)--(1);
\end{tikzpicture}
\caption{Cycle 4}
\end{subfigure}
\centering
\begin{subfigure}{0.32\textwidth}
\centering
\begin{tikzpicture}[scale=0.85, every node/.style={circle,draw,inner sep=2pt}]
    \node (1) at (90:2) {1}; \node (2) at (30:2) {2};
    \node (3) at (-30:2) {3}; \node (4) at (-90:2) {4};
    \node (5) at (-150:2) {5}; \node (6) at (150:2) {6};
    \draw[copytwo] (1)--(4)--(5)--(3)--(6)--(2)--(1);
\end{tikzpicture}
\caption{Cycle 5}
\end{subfigure}

\caption{Hamiltonian cycle packing for $K_6$. The edges originating from the two distinct copies are respectively coloured red and blue.}
\label{fig:K6-final}
\end{figure}

\typea*
\begin{IEEEproof}
Observe that $A=\cM$, and let $q=m-1$. We break the proof into the following pair of cases.

 \textbf{Case 1:} $m$ is even, that is $q$ is odd. 
 
 In this case, we show that for every $i\in\cM$, a single copy of $P_{K_{m,3},i}$ admits a Hamiltonian decomposition into $\frac{q-1}{2}$ cycles. Hence, the edge reuse number for every hyperedge is $t_e=\sum_{i\in\cM}\mathbf{1}\{i\in e\}=3$. Thus, using a blocklength of $n=3$, Theorem~\ref{th:3unif} ensures that a perfect secret key of rate $\frac{2}{3}\sum_{i\in\cM}\frac{q-1}{2}=\frac{m(m-2)}{3}$ can be generated. Lemma~\ref{lem:typescomp} thus implies that this perfect secret key achieves capacity. 

 It remains to show that $P_{K_{m,3},i}$ admits a Hamiltonian decomposition into $\frac{q-1}{2}$ cycles. Firstly, observe that $P_{K_{m,3},i}$ is nothing but the complete graph on $m-1=q$ vertices. Since $q$ is odd, we invoke Walecki's Theorem \cite{alspach2008wonderful}, Theorem 5.2, which states that the complete graph on an odd number (say $q$) of vertices admits a Hamiltonian decomposition of size $\frac{q-1}{2}$. This completes the proof of Case 1.

\textbf{Case 2:} $m$ is odd, that is $q$ is even.

 In this case, we show that for every $i\in\cM$, two copies of $P_{K_{m,3},i}$ admit a Hamiltonian decomposition into $q-1$ cycles. Hence, the edge reuse number for every hyperedge is $t_e=\sum_{i\in\cM}2\mathbf{1}\{i\in e\}=6$. Thus, using a blocklength of $n=6$, Theorem~\ref{th:3unif} ensures that a perfect secret key of rate $\frac{2}{6}\sum_{i\in\cM}(q-1)=\frac{m(m-2)}{3}$ can be generated. Lemma~\ref{lem:typescomp} thus implies that this perfect secret key achieves capacity. 

 It remains to show that two copies of $P_{K_{m,3},i}$ admit a Hamiltonian decomposition into $q-1$ cycles. Recall that $P_{K_{m_3},i}$ is essentially the complete graph $K_q$ on $m-1=q$ vertices. For $q \in \{4, 6\}$, we provide the explicit decompositions in Figures~\ref{fig:K4-final} and~\ref{fig:K6-final}. For $q\geq 8$, we make use of a result by Tillson~\cite{Tillson1980AHD}, which states that the complete directed hypergraph on $q$ admits a decomposition into $q-1$ directed Hamiltonian cycles, provided $q\geq 8$. Now, a complete directed graph can be viewed as two copies of the complete graph with the orientations of the edges being discarded. Thus, Tillson's theorem provides a Hamiltonian decomposition of two copies of $K_q$ into $q-1$ Hamiltonian cycles as required. 
\end{IEEEproof}

\typeb*
\begin{IEEEproof}
Choose $A=\{i\}$, and recall that the hypothesis of Corollary~\ref{lem:generic_hamdecomp} states $P_{\cH,i}$ admits a Hamiltonian decomposition into say $p$ cycles. Thus, edge reuse number $t_e=1$ for every $e\in\cE$, and hence choose $n=1$. Therefore, the scheme of Theorem~\ref{th:3unif} generates a perfect secret key of rate $2p$. 

Now, $P_{\cH,i}$ being a graph on $m-1$ vertices, each of the Hamiltonian cycles contains $m-1$ edges. Next, note that if a graph admits a Hamiltonian decomposition, then each edge must participate in exactly one of the cycles. Therefore, we must have $|\cE|=|\cE_i|=p(m-1)$. Therefore,
 we have by \eqref{eq:typeS} that
 $$\csk(X_{\cM})\leq I_{\cS}(X_{\cM})=\frac{2}{m-1}p(m-1)=2p\leq \csk(X_{\cM}),$$ where the last inequality follows by noting that Theorem~\ref{th:3unif} ensures a key of rate $2p$. Thus, $\csk(X_{\cM})=I_\cS(X_\cM)=2p$, and hence the source is Type-$\cS$. Also, the perfect secret keys generated by Theorem~\ref{th:3unif} achieve capacity.

\end{IEEEproof}

\typec*
\begin{IEEEproof}
Choose $A=V_1$. We first claim that for any $i\in V_1$, the graph $P_{\cH,i}$ is the complete bipartite graph $K_{2r,2r}$. This follows by noting every edge in $P_{\cH,i}$ is of the form $\{j_1,j_2\}$ where $j_1\in V_1\setminus\{i\}$ and $j_2\in V_2$. Hence, $P_{\cH,i}$ is bipartite with two parts $V_1\setminus\{i\}$ and $V_2$, and $|V_1\setminus\{i\}|=|V_2|=2r$. Furthermore, every subset $\{j_1,j_2\}$ with $j_1\in V_1\setminus\{i\}$ and $j_2\in V_2$ is an edge in $P_{\cH,i}$, thanks to the hyperedge $\{i,j_1,j_2\}\in\cE$. Thus, $P_{\cH,i}$ is indeed $K_{2r,2r}$. 

It is a standard result of graph theory that a single copy of $K_{2r,2r}$ admits a Hamiltonian decomposition of size $r$ -- see, for example, the Introduction section of \cite{LA}. Thus, for every $i\in A$, we have $k_i=1$ and $p_i=r$. Furthermore, note that for every $e\in\cE$ the edge reuse number $t_e=2$. This follows by noting that any hyperedge $\{j_1,j_2,j_3\}$, with $j_1,j_2\in V_1$ and $j_3\in V_2$, is used twice, once in $P_{\cH,j_1}$ and once in $P_{\cH,j_2}$. Then, invoking the scheme of Theorem~\ref{th:3unif} with a blocklength of $n=2$ gives us a perfect secret key of rate $\sum_{i\in A}r=r(2r+1)$.

Finally, we have using \eqref{eq:typeS} that
$$
\csk(X_{\cM})\leq I_\cS(X_\cM)=\frac{2}{4r}|\cE|=\frac{1}{2r}\frac{2r(2r+1)}{2}2r=r(2r+1)\leq\csk(X_\cM),
$$
where the last inequality follows by noting that Theorem~\ref{th:3unif} gives a perfect secret key of rate $r(2r+1)$. Therefore, this perfect secret key achieves capacity, and the source is Type-$\cS$.

\end{IEEEproof}

\end{appendices}

\bibliographystyle{ieeetr}
\bibliography{references}

\end{document}